# State sharding model on the blockchain

Xiangyu Wang, Ting Yang, Yu Wang

## Abstract

Blockchain is an incrementally updated ledger maintained by distributed nodes rather than centralized organizations. The current blockchain technology faces scalability issues, which include two aspects: low transaction throughput and high storage capacity costs. This paper studies the blockchain structure based on state sharding technology, and mainly solves the problem of non-scalability of block chain storage. This paper designs and implements the blockchain state sharding scheme, proposes a specific state sharding data structure and algorithm implementation, and realizes a complete blockchain structure so that the blockchain has the advantages of high throughput, processing a large number of transactions and saving storage costs. Experimental results show that a blockchain network with more than 100,000 nodes can be divided into 1024 shards. A blockchain network with this structure can process 500,000 transactions in about 5 seconds. If the consensus time of the blockchain is about 10 seconds, and the block generation time of the blockchain system of the sharding mechanism is 15 seconds, the transaction throughput can reach 33,000 tx/sec. Experimental results show that the throughput of the proposed protocol increases with the increase of the network node size. This confirms the scalability of the blockchain structure based on sharding technology.

## 1. Introduction

In November 2008, the article "Bitcoin: Peer-to-Peer Electronic Cash System" published by Satoshi Nakamoto [1] published the theoretical basis of blockchain technology. In January 2009, the Bitcoin system was born, and it was the first application based on blockchain technology. The Bitcoin system is a payment system used to conduct online transactions directly from one party to another without trusting a third party. In December 2013, the Ethereum blockchain (Ethereum) [2] platform based on blockchain technology was born. Ethereum can not only support the transaction of digital currency Ether, but also provides a Turing complete programming language to write smart contracts. In December 2015, the Linux Foundation initiated the Hyperledger [3] open project, which aims to promote the application of blockchain technology across industries. The main characteristics of blockchain technology include decentralization, openness, transparency, and gradual, but the overall blockchain architecture faces the problem of poor scalability. Zhu Yanjie [4] and others mentioned mainly include two aspects: storage Capacity issues, transaction blasting issues. In August 2020, the 2020 Global Blockchain Innovation and Development Conference with the theme of "Connecting Value and Linking the World" was held, which put forward more requirements on blockchain technology.

Sharding technology is a technology to improve the scalability of the blockchain. The blockchain protocol based on sharding technology alleviates the problem of node scale, latency and low throughput in traditional blockchain protocols. The sharding technology refers to Internet database technology [5], and the core idea of the blockchain sharding technology is to divide the entire blockchain network into multiple sub-networks. Each sub-network is a shard, and each shard is for a different A subset of transactions are processed, verified and reached consensus. All shards in the network process different transactions in the network in parallel, and there is no need for frequent communication between each shard, thereby improving the performance of the blockchain. The sharding in the blockchain is divided into network sharding, transaction sharding and state sharding according to the object [6].

State sharding stories different states in different shards to reduce the storage burden of each node, including account information or smart contract information [7]. It can reduce the redundant storage of state, making the entire blockchain network have storage scalability. Unfortunately, there are almost no research on state sharding, mainly due to two challenges. The first challenge of state sharding is frequent cross-shard communication and state exchange. Since each shard only stores part of the state of the system, the communication between the shards is necessary to obtain the information stored in other shards. Information, the only way to verify whether the transaction is valid. The second challenge is data availability. When a node or shard fails, transactions from this node or shard will not be verified, which will cause the entire system to crash.

In this article, we propose a new blockchain protocol called SSChain, which aims to solve the storage scalability problem of blockchains by combining state sharding technology. The key idea of SSChain is to use distributed hash table (DHT) technology to create random shards, and then use the Merkle Patricia Tree (MPT) data structure to store the mapping from the blockchain account address to the account state, and the account state Using Merkle Directed Acyclic Graph (Merkle DAG) data structure, finally, all account state data is stored in DHT.

The rest of this article is organized as follows. Section 2 introduces related work, section 3 introduces the model, section 4 introduces the design of SSChain blockchain structure, section 5 introduces the implementation of SSChain algorithm, and section 6 introduces our simulation experiment results. The conclusion is drawn in the last section.

## 2. Related work

In recent years, on one hand, as the block chain has become larger and larger (making it difficult to store, send, receive and manage), Liu Xizi [6] analyzed the current status of global block chain technology and application innovation, and the sharding technology innovation plan will Breaking through the blockchain performance bottleneck and sharding technology is also an important driving force for the development of blockchain technology. The sharding technology is borrowed from Internet database technology [5]. The core idea of the blockchain sharding technology is to divide the entire blockchain network into multiple sub-networks. Each sub-network is a shard, and each shard is for a different A subset of transactions are processed, verified and reached consensus. All shards in the network process different transactions in the network in parallel, and there is no need for frequent communication between each shard, thereby improving the performance of the blockchain.

On the other hand, some researchers have proposed some solutions to reduce the size of the ledger in the blockchain system. For example, in the 2018 CVCBT conference, three authors Zhijie Ren [8], Emanuel Palm [9] and Michal Zima [10] respectively proposed a VTL system to trim blockchain transactions and reduce redundant data waste in transaction input. To reduce the size of the ledger in the blockchain system; among them, the sharding method proposed by Zhijie Ren relies on the VTL system, and the nodes can only be sharded under a few specific conditions, which cannot popularize all blockchain systems; secondly, VTL The system cannot guarantee that every node in the shard will behave rationally. The definition of the predicate function proposed by Emanuel Palm is only a semantic and simple function implementation, and does not consider complex trading conditions. To expand, the function just judges the transactions of interest and deletes the size of the ledger, and some transactions will not be traced. In general, the current research on blockchain state sharding technology is too limited to solve the storage scalability of blockchain. Therefore, the research on blockchain state sharding technology is necessary. Valuable, the problem studied in this article is to support high-throughput blockchain state sharding technology research.

## 3. Model

The ordinary distributed hash table (DHT) builds its topology based on the structure diagram. For most hash tables, the following principles are true: each node of the system will be assigned a unique identifier, and the identifier space (for example, 256 bits) String set) Partition among all nodes of the system. The nodes are self-organized in the graph according to the distance function based on the identifier space.

The concept of Shared DHT is similar to ordinary DHT, except that each vertex of shared DHT is a group of nodes instead of a single node. Each group of nodes has a unique common prefix, and the prefix number can be used as a label for the shard. That is, nodes are clustered together to form fragments, and the fragments self-organize into a DHT graph topology.

Assuming that each node in the blockchain has saved its own state, in the blockchain state sharding, we perform the same Hash algorithm calculation on the node (host) and account or smart contract address and map the result to the Hash space in. The nodes in the Hash space only save the account state, that is, a copy of the account data. In this way, each node in the network does not need to save the complete blockchain state, and only needs to save a hash relationship mapping table to obtain all the data of the blockchain. State fragmentation reduces the pressure on nodes to store data and releases unnecessary space resources. The basic model diagram of state fragmentation is shown in Figure 1. First use the same Hash function to map the node and data to the Hash space, and then map the data to the node. You can select the IP address or account address of the node for hash calculation and map it to the Hash space (mod $2^{256}$). Use the same hash algorithm to calculate the hash value of the key of the data, such as the hash of the transaction, and map the position of the data on the hash ring. The position moves clockwise along the ring, and stores the data in the first encounter node.

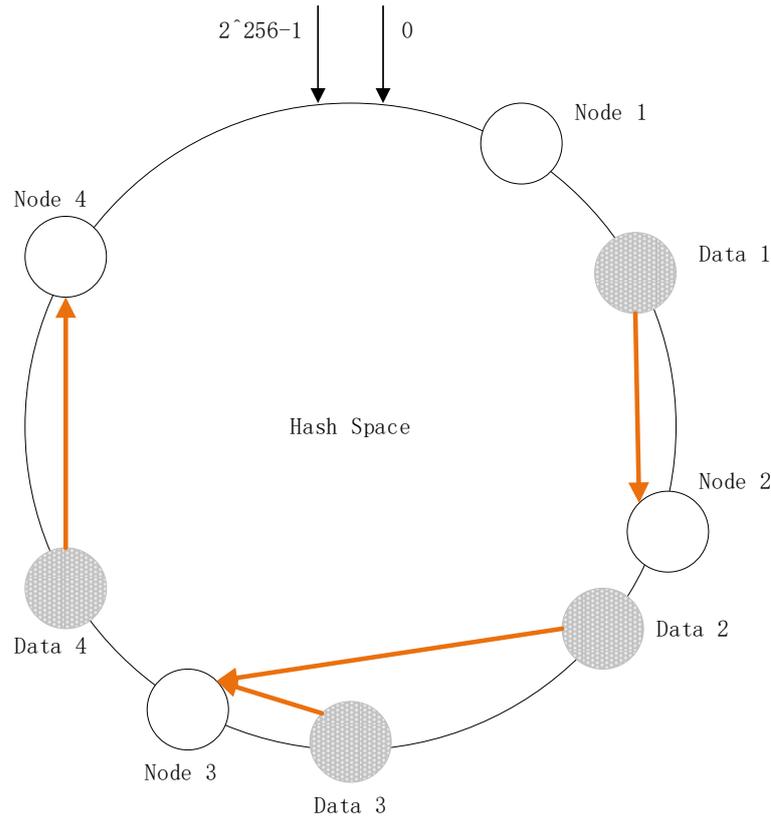

Figure 1 Basic model of state fragmentation

## 4. Design of SSChain blockchain structure
### 4.1 Storage structure

  The current blockchain is to ensure that each node maintains the entire "ledger". The status data of the account is ultimately stored in the database in the form of key-value pairs. Due to the continuous increase of the blockchain ledger, the storage capacity requirements of the nodes are more stringent. In order to save storage costs for nodes participating in the maintenance of the blockchain system, this article proposes to improve the existing blockchain structure. The SSChain block chain structure proposed in this article is based on a combination of Merkle prefix tree (Merkle Patricia Tree, MPT) and Merkle Directed Acyclic Graph (Merkle Directed Acyclic Graph, Merkle DAG) data structure, and uses distributed hashing the storage structure of the table (Distributed Hash Table, DHT) stores the SSChain blockchain.

  On contrast, the SSChain blockchain structure proposed in this article uses MPT to store the mapping from the blockchain account address to the account status. The account status is divided into two types: external account status and contract account status. Account status generally includes account balance and account transaction number. For contract accounts, Including contract code and contract storage. The account status designed in this paper adopts the Merkle DAG data structure. With the passage of time, the historical status of the account gradually increased, which eventually caused the storage space requirements of the node to gradually increase. If using ordinary storage management methods to save data, this will consume huge storage resources, but using a distributed storage system to manage the state of all accounts in the blockchain will greatly save storage space.

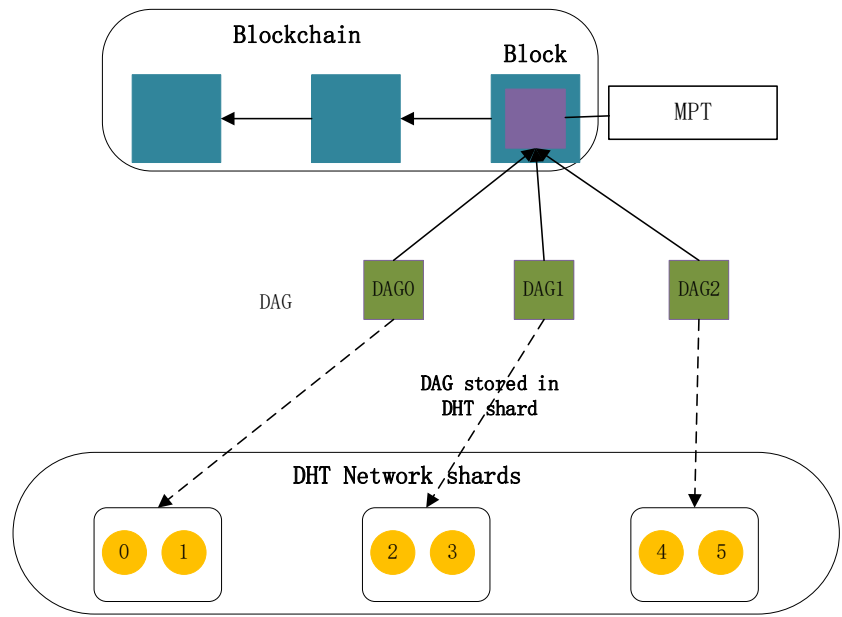

Figure 2 Architecture of SSChain

As Figure 2 shows, to store DAG structure in DHT network shards, we define the structure. First of all, each DAG structure is a leaf node of the MPT number, indicating the status of the account. Then, each account has a unique ID, and each DHT network segment is also identified by a unique segment number. According to Figure 1, each account status and segment are mapped to the DHT ring. Finally, according to the mapping relationship between the account ID and the shard number, the account status is stored in the shard.

Function definition of each noun in picture 2 are as follows.

1. Merkle Patricia Tree (MPT): The MPT data structure is a key-value storage structure. The encoded account address is used as the key value of MPT, and the account state encapsulated by the Merkle DAG data structure is used as the value value. In general, the key value of the MPT represents the search path of the MPT tree (the path stores the encoded account address), and the value value represents the leaf node of the MPT tree (the leaf node stores the account status). The data encapsulated in the Merkle DAG structure is suitable for storage in a distributed storage system, so this article uses a distributed storage system based on the DHT structure to store the account state.

2. Merkle Directed Acyclic Graph (Merkle DAG): The Merkle DAG data structure is a data structure suitable for distributed storage of files. Merkle DAG has the following functions: 1) Content addressing: Use multiple hashes to uniquely identify the content of a data block; 2) Anti-tampering: It is convenient to check the hash value to confirm whether the data has been tampered with; 3) Deduplication: due to the hash of data blocks with the same content is the same, which can easily remove duplicate data and save storage space. If you store all the states of each account in the blockchain in one file, and then use the Merkle DAG data structure to manage the version information of each file, this method will save storage overhead.

3. Distributed Hash Table (DHT): DHT is a distributed storage method. Without

the need for a server, each client is responsible for a small range of routing and for storing a small part of data, thus realizing the addressing and storage of the entire DHT network. DHT serves as the underlying storage network architecture of the blockchain, and the account status data encapsulated with the Merkle DAG data structure will be stored in a specific node group in the DHT network.

## 4.2 State Update

In the current blockchain system, such as Ethereum, if the state of an account in the blockchain system changes, each full node needs to reconstruct an MPT tree locally. However, there are a small number of accounts whose state changes in the blockchain, and not all account states change at the same time, so the update of the account state is a partial update. Because the MPT structure has the advantage of partial update, this article chooses MPT as the storage structure of the blockchain account address to account state mapping. Because the Merkle DAG structure supports block rollback or account state rollback, and is perfectly compatible with distributed storage technology, this article uses Merkle DAG to store the blockchain account state. If all full nodes construct an MPT tree locally, this will waste huge computing resources and storage resources. Therefore, a distributed storage system, such as IPFS, is used to store the account status or data in the blockchain, which can greatly save the storage space of the node.

This article chooses the MPT data structure to store the mapping relationship between the account address and the account state, and the account state uses the Merkle DAG structure. The account data is finally stored in the database in the form of key-value pairs. This article uses DHT distributed technology to store account data. The MPT data structure has the characteristics of high query efficiency and excellent local status update performance. The query time is constant and complex. Since most of the account states in the blockchain system are unchanged, only a small number of accounts need to be updated and only need to be accessed. The corresponding branch can modify the account status. It is more convenient to write the corresponding status or data of the account into a file and use the Merkle DAG data structure to manage the account data status in a version controlled manner. DHT is the core technology of the distributed system. According to the core idea of DHT distributed storage, different account addresses have different attributes and are mapped to different node groups. The account state is stored in the node group to which the address is mapped, which completes the blockchain. State sharding enables different account states to be stored in different shards (node groups). Using DHT distributed storage technology to reduce the redundant storage of the account state, so that the entire blockchain has storage scalability.

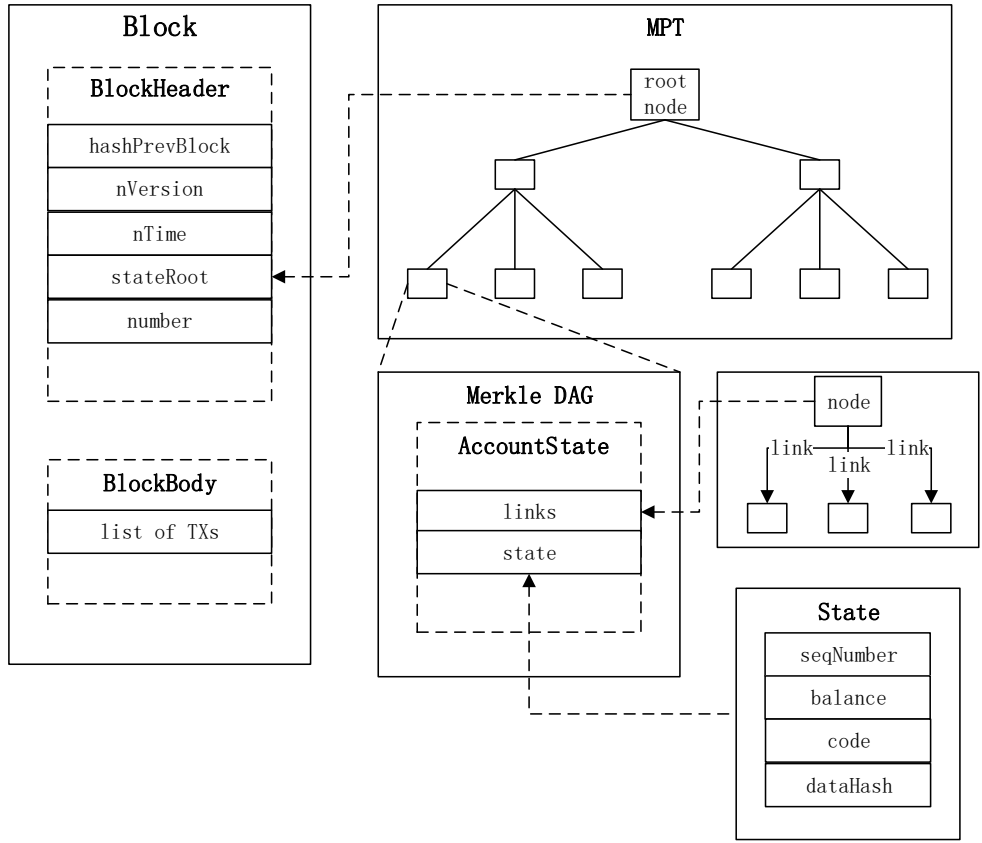

Figure 3 Block structure of SSChain

The block structure of SSChain blockchain is shown in Figure 3. As the Figure 3shows, the block of SSChain consists of two parts, namely the BlockHeader and the BlockBody, which are almost the same as the traditional blockchain structure, except that one more stateRoot is added. Field. Therefore, this article only describes the stateRoot field in the block header. StateRoot is the hash of the state, which is the hash value after RLP encoding. The value is derived from the root calculated by the MPT tree structure. For the leaf nodes of the MPT tree structure, the leaf nodes store the data of the account state. The state has gone through a layer of encapsulation, that is, encapsulated by the Merkle DAG structure. The data is stored in the DHT distributed hash table, which is similar to the version management method to manage the account state or data.

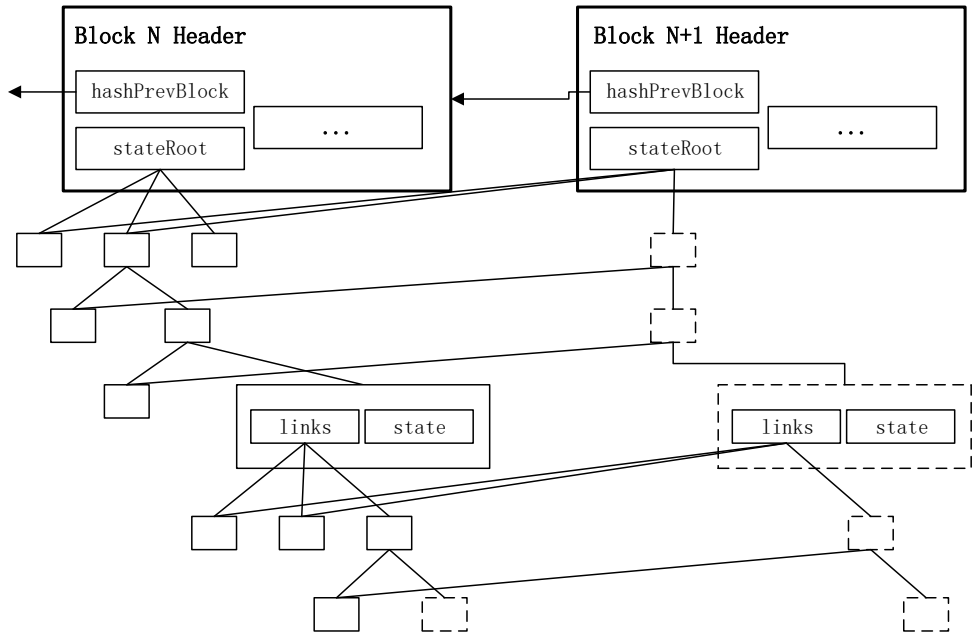

Figure 4 Blockchain structure of SSChain

The blockchain structure is shown in Figure 4. As shown in Figure 4, every time a new block is generated, the state of the SSChain blockchain will not modify the original MPT tree, but will create some new branches. When the status of an account changes, the original branch node data of the MPT tree will not change. New MPT tree structure can be constructed and new stateRoot values can be generated, inserted into the block. Farther, because the nodes in the blockchain maintain the MPT tree containing the Merkle DAG structure, each time a block modifies the state of the MPT tree is only a small part, that means minor modification of the Merkle DAG structure. Therefore, the structure is not only conducive to block rollback or account state rollback, but also reduces redundant storage of account status.

## 5. SSChain algorithm implementation

This chapter will introduce the SSChain algorithm in detail. SSChain uses the MPT data structure as the data structure of the mapping relationship between the account address and the account state, and the account state (data) uses the distributed Merkle DAG data structure for version management.

SSChain terms are defined as follows:

The function to be completed by the MPT data structure is: the mapping of account address (key) to account status (value). The account address used by Ethereum is 160 bits, which is generally expressed as 40 hexadecimal numbers. The status is the status and status of the external account. The status of the contract account, including the balance, the number of transactions, and for the contract account, it also includes the code and storage.

The key encoding in the MPT data structure includes three encoding methods: Row encoding, Hex encoding and HP (Hex Prefix[11]) encoding, and RLP (Recursive Length Prefix[12]) encoding is used for [key, value]. The MPT data structure takes [key, value] as v after RLP encoding, calculates the hash of the encoded data as k, and stores it in levelDB (levelDB is a k-v database).

Row encoding: It is the original byte order array encoding method.

Hex encoding: When [k,v] data is inserted into MPT, their k(key) must be encoded. At this time, for the key encoding, it is necessary to ensure that the key originally of the byte array type can enter the child nodes of the MPT tree bit by bit in hexadecimal form (regulation), which solves the problem of loading the node key value into memory.

HP encoding: Its main purpose is to restore the Hex format string to the byte format, that is, to put the node into the disk for permanent storage.

RLP encoding: RLP encoding is also known as RLP serialization. It is a simple and efficient encoding method. It mainly uses the same encoding function to recursively encode in the forward direction until it can no longer be encoded. The same process is also applicable to decoding, it is only necessary to substitute the decoding results into the same decoding function until it is completely plaintext.

The relationship between the four codes is shown in Figure 5.

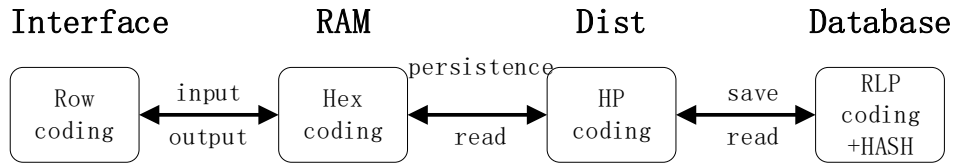

Figure 5 Four coding relationships

Step 1 SSChain algorithm steps

Step 1.1 Suppose that the "accounting" node N has the state sharding authority, and node N first performs Hex encoding on the address and state (or data) of account A to be saved.

$$Book_N = 1 \cap Authority_N = 1 \Rightarrow \left( Address_A' = Hex(Address_A) \cap State_A' = Hex(State_A) \right)$$

Step 1.2 Node N performs HP coding, RLP coding and HASH calculation on the address after the account A is coded, and then performs query operations on the obtained results in the DHT distributed database system.

$$Y = Inquire\left( Hash\left( RLP\left( HP\left( Address_A' \right) \right) \right) \right) = [k, v]$$

$$\left( N = Inquire \cap return(Y) \right) \cup \begin{pmatrix} N = Update \cap Y = \left[ K, Root(generate(MerkleDAG)) \right] \\ \cap \operatorname{Re}turn\left( Root(generate(MPT)) \right) \end{pmatrix}$$

step 1.3 Suppose the result of the query is Y=[k,v]. If the node N is a query operation, the query result Y is returned and ends; if the node N is an update operation, the MPT tree and Merkle DAG structure are regenerated, and the The value of v in Y is replaced with the root of the Merkle DAG structure, which returns the root of the MPT tree.

step 1.4 Update the status of account A in the DHT distributed database, and the status segment update operation ends.

$$State_A' = Update(State_A)$$

Founction 1 is the implementation of related functions of the state fragmentation algorithm, andpicture 6 is the overall flow chart of the implementation of the state fragmentation algorithm. The time complexity of the entire state slicing algorithm depends on the complexity of constructing the MPT tree structure and the Merkle DAG structure algorithm, which is O(n) (here, n SHA256 operations).

Founction 1 Code of SSChain

**Algorithm 1 SSChain algorithm**

**Input:** address: address of a node or account; data: data of the account; DHT: DHT database; rootHash: state root hash of the MPT tree

**Output:** stateRoot: state root of the MPT tree

**Procedure** Init()   **//**initialize MPT tree

    if(DHT == empty) then

        exit();

    end if

    if(rootHash == empty) then

        search DHT database;

    end if

    if(DHT != emtpy && rootHash != empty) then

        MPT = createMPT(rootHash); //create MPT tree and return it's root hash

    end if

    update DHT database;

    update MPT tree and Merkle DAG;

**End Procedure**

**Procedure** createMPT() // Create MPT tree and return it

    hashNode = DHT.bytesToHash(rootHash);//data of RLP from DHT

    decodeNode(hashNode);//decode the data

    return MPT;

**End Procedure**

**Procedure** insert()   //insert data to MPT tree and DHT database

    if (len(key) == 0) then

        return null;

    end if

    judege types of the node;

    insert();

**End Procedure**

**Procedure** Traversal () //Traversal a MPT tree and get the value of the key

```
        if (len(key) == 0) then
                return null;
        end if
    judege types of the node;
    if(key is valid) then
            resolveHash(value);
            return value;
    end if
    Traversal ();
End Procedure
```

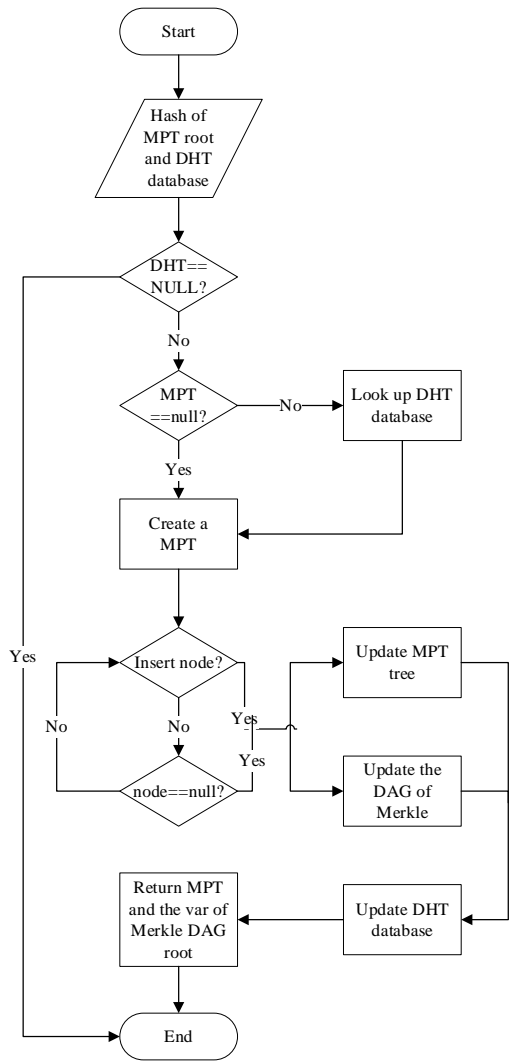

Figure 6 SSChain algorithm flow chart

Figure 6 is the overall flow chart of SSChain algorithm implementation. First,

the system calls the initialization function. The function accepts a hash value and a distributed database parameter. If the hash value is not a null value, it means that an existing MPT tree is loaded from the database, and the resolveHash method is called to load the entire MPT tree. If rootHash is empty, then create a new MPT tree and return. Creating an MPT tree is a recursive operation. The insert method is called recursively, starting from the root node, and looking down until it finds a point that can be inserted, and insert it. The type of node inserted into the MPT tree requires the key value to undergo Hex encoding and RLP encoding, while the value value is stored in the Merkle DAG structure. The value value also needs RLP encoding and HASH calculation. Finally, the root hash value of the MPT tree is obtained. After the root hash value of Merkle DAG and Merkle DAG, it needs to be stored in the DHT distributed database, that is, the data state of the blockchain is saved through version management.

## 6. Experiment

The experimental environment parameters used in this section are shown in Table 6-1.

Table 1 Experimental environment parameters

| Hardware | PC：Inter® Core™ i7-6500 CPU 2.50GHz（4 CPUs） 内存 12GB |
|---|---|
| Software | Windows 7 64bit 和 go-ipfs v0.4.23 |

This chapter mainly conducts experiments on Merkle DAG structure and DHT distributed storage technology. The software used in the experiment is IPFS. First, I conduct an experiment on the distributed network to check the information of routing nodes in the DHT distribution, as shown in Figure 6-1.

Figure 6-1 Node information in the IPFS network

Use the "ipfs swarm peers" command to view the nodes that are using the IPFS network near the node, which also means that the IPFS network is a DHT distributed network. I downloaded some block data (such as Ethereum) from the blockchain network and made some parameter modifications. As shown in Figure 6-2, I downloaded the status (data) files of four users and used the "blocks" folder Save these four files.

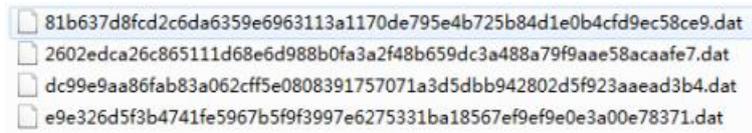

Figure 6-2 Four user status (data) files

Next, I use the "ipfs add -r" command to upload the entire blocks directory. The upload result is shown in Figure 6-3.

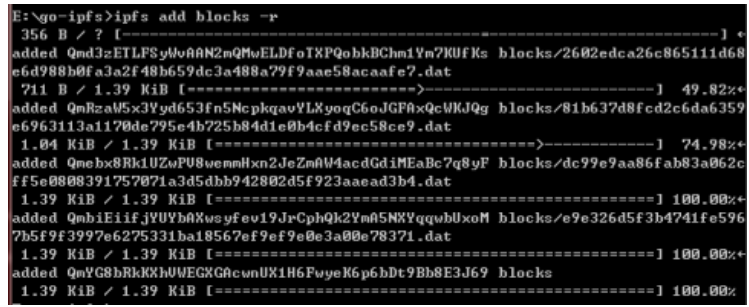

Figure 6-3 Upload user status (data) file

Uploading files to the IPFS network is performed recursively. The sub-files in the directory are uploaded first, and the directory files are uploaded last, and the data in each file is encoded and hashed to prevent tampering and comparison of information.

When we upload the block file to the IPFS network, it will return a Merkle DAG root hash value to us. We can use this value to get the contents of the entire folder in the IPFS network, as shown in Figure 6-4.

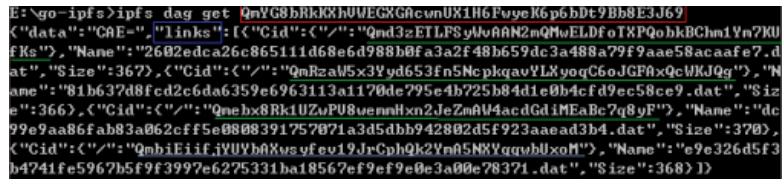

Figure 6-4 View the Merkle DAG structure of the directory

Through the "ipfs dag get" command, we can view all the information of the entire folder, where "links" means all link information in the directory, and here means there are four files in the blocks directory. Through the corresponding hash value, we can use the "ipfs cat" command to get the data in the file, that is, the account status. As shown in Figure 6-5.

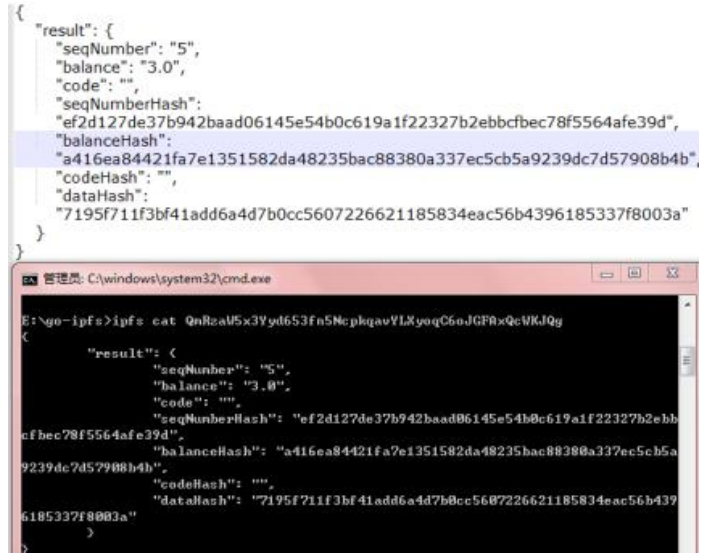

Figure 6-5 Get account status (data)

There is a domain name space in IPFS, which can dynamically change data in an IPFS distributed network. When we modify the state of the account in the block, we can update the data to the IPFS domain name space, allowing our node's domain name space to reference another IPFS hash, that is, we can use the node ID to perform the hash value After binding, we can directly access the block information through the node ID in the future. When we update the block information or account status, we can republish it to the IPFS domain name space.

Figure 6-6 shows using the "ipfs name publish" command to upload to the DHT distributed network for storage, and return the most recently stored address, which is the node ID.

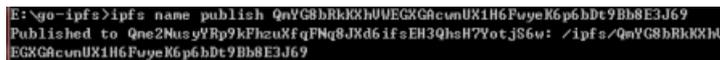

Figure 6-6 Publish files to IPFS domain name space

Since this article stores the state of the account in the Merkle DAG structure, that is, in the IPFS network, when the account in the blockchain performs transactions, the state of the account will also change, that is, the state file information stored in the IPFS network of the corresponding account will also be Update, and the hash value on the corresponding Merkle DAG path will also change. Figure 6-7 shows that two of the account statuses are updated (the second and third underlines are indicated), the "data" field in the Merkle DAG structure is also automatically updated, and the hash value on the path will also change, which eventually leads to The root hash value of Merkle DAG is updated (represented by the first rectangle).

Figure 6-7 Update account status (data)

After we obtain the updated Merkle DAG root hash value, we can view the number of updated account files by using the "ipfs cat" command. As shown in Figure 6-8.

Figure 6-8 Account status (data) details after update

Overall, this chapter conducts experiments on the blockchain that includes a state sharding mechanism. The experiment uses the IPFS distributed file system to complete the storage and update of the account state.

# 7. Conclusion

This paper designs and implements the state sharding scheme of the blockchain network. State sharding is the process of storing different states or data of an account in different shards to reduce the storage burden of each node. This chapter proposes to use the MPT data structure to save the state of the account in the state shard, and the Merkle DAG data structure to save the historical version information of the account state in the state shard, to facilitate the rollback and update of the state, and the use of DHT distributed storage technology to reduce the account state Redundant storage. This paper proposes a state sharding algorithm and implements a complete blockchain structure including state sharding mechanism.

## REFFRENCES

[1] S. Nakamoto. Bitcoin: A Peer-to-Peer Electronic CashSystem [EB/OL]. https://bitcoin.org/bitcoin.pdf, November 7, 2019.

[2] Ethereum Community Authors. A Next-Generation Smart Contract and Decentralized Application Platform [EB/OL]. https://github.com/ethereum/wiki/wiki/White-Paper, April 19, 2019.

[3] E. Androulaki, A. Barger, V. Bortnikov, et al. Hyperledger fabric: a distributed operating system for permissioned blockchains[C]. Thirteenth EuroSys Conference. ACM, 2018: 1-15.

[4] Zhu Yanjie, Zhang Zhisheng, Duan Lin. Scalability analysis of mainstream open source blockchain framework[J]. Yunnan Electric Power Technology, 2018, 46(06): 26-28.

[5] Liu Jingyi. Exploring database sharding technology[J]. Network Security and Information Technology, 2017(06): 81-82.

[6] Liu Xizi. Global blockchain technology and application innovation status, trends and enlightenment[J]. Science & Technology China, 2020(1): 27-31.

[7] Y. Gao, S. Kawai, H. Nobuhara. Scalable Blockchain Protocol Based on Proof of Stakeand Sharding[J]. Journal of Advanced Computational Intelligence and Intelligent Informatics. 2019, 23(5): 856-863.

[8] Z. Ren, K. Cong, T. Aerts, et al. A Scale-Out Blockchain for Value Transfer with Spontaneous Sharding[C]. 2018 Crypto Valley Conference on Blockchain Technology (CVCBT), Zug, 2018, 1-10.

[9] E. Palm, O. Schelén and U. Bodin. Selective Blockchain Transaction Pruning and State Derivability[C]. 2018 Crypto Valley Conference on Blockchain Technology

(CVCBT), Zug, 2018, 31-40.

[10] M. Zima. Inputs Reduction for More Space in Bitcoin Blocks[C]. 2018 Crypto Valley Conference on Blockchain Technology (CVCBT), Zug, 2018, 112-115.

[11] Ethereum Community Authors. Hex Prefix Encoding [EB/OL]. https://github.com/exthereum/hex_prefix, February 24, 2020.

[12] Ethereum Community Authors. Recursive Length Prefix Encoding [EB/OL]. https://github.com/ethereum/wiki/wiki/RLP, December 9, 2019.